\documentstyle[12pt]{article}

\begin{document}
\vspace{0.1in}
\begin{center}
{\Large  \bf  Optical soliton solutions
of the quintic complex Swift-Hohenberg equation\\}
\vspace{0.3cm} {\large    \underline{  A. Ankiewicz}$^1$, K. Maruno$^2$ and
N. N.
Akhmediev$^3$
}
  \\ \vspace{0.1in}
$^1$ Applied Photonics Group, and $^3$ Optical Sciences
Centre, Research School of Physical Sciences and Engineering,
\\ The Australian National University, Canberra ACT 0200,
Australia \\ $^2$  Research Institute for Applied Mechanics, Kyushu University,Kasuga, Fukuoka, Japan.

\end{center}

\vspace{0.2cm}
{\bf 1. Introduction.}
 Complicated optical systems with gain and loss can be described by the
complex Swift-Hohenberg (S-H) equation. 
These include synchronously pumped optical
parametric oscillators, pattern formation in convection cells and passively
mode-locked lasers with special spectral features \cite{r4}-\cite{Weiss}.
The main difference
between the complex S-H equation and previously-studied master 
equations lies in
its more
involved spectral filtering term. At the same time, these complications do
not allow us to analyse the solutions easily. In fact, it was not
clear that such solutions may exist at all \cite{r4}-\cite{Weiss}. In this
work, we study the
quintic complex S-H equation in 1D and report various new exact solutions.

The normalized quintic complex S-H equation is\cite{SB98}:
\[
i\psi_{z  }+\frac{D}{2}\psi_{tt}+|\psi|^2 \,\psi+\,(h+i s)\psi_{tttt}
+ (\nu\,-i \mu) |\psi|^4 \psi=i\delta \psi
+ i \beta \psi_{tt} + i \epsilon |\psi|^2 \psi.
\]
In mode-locked laser applications, $z$ is the propagation distance
or the cavity round-trip number (treated as a continuous variable), $t$ is
the retarded time, $D$
is the 2nd order dispersion, $h$ is the 4th order dispersion,
$\epsilon$ is a
nonlinear gain (or 2-photon absorption if negative) and
$\delta$ (usu. negative) represents a constant gain or loss. The
band-limited gain (e.g. due to an EDFA, where the gain band may be
about 30 nm around 1.5 microns) is represented by $\beta$
(parabolic spectrum shape) and $s$ (4th order correction).
We find exact forms for a range of solitons, including bright
and dark cases, and both chirped and unchirped forms. They have
various features which differentiate them from solitons of the
complex Ginzburg-Landau equation \cite{book}. These first solutions
can give some clues for analysing more involved solutions.

{\bf 2. Analysis.}
Let $\psi=f(\,t  )\,exp[ -i\,\, \Omega\,z\, ]$.
This reduces the quintic complex S-H equation to an ordinary differential eqn. in $f$:
\begin{equation}\label{A4}
 \Omega   \,f   +(\frac{D}{2}-i
\beta)\,f''(\,t  )+(1-i\,\epsilon)|f|^2 \,f-i\delta f+\,(h+i
s)\,f''''(\,t  )  + (\nu\,-i \mu) |f|^4 \,f=  0.
\end{equation}
We will show that basic [unchirped]
 bright and dark solitons exist, and that there are also chirped
bright and dark solitons.

{\bf 3.1 Bright soliton}
If we take $f=c\,g\, sech(g\,\,t  )$, then we note that
$
 f''(t  )/[g^2\,f(t  )]  = \, \, 2 \,tanh^2(g\,t  )-1
$
and that
$
 f''''(t  )/[g^4\,f(t  )] = \, \, 5-28\, tanh^2(g\,t  )+24
\,tanh^4(g\,t  ) .
$
On dividing through by $f$, it is clear that the coefficients of
$tanh^n(gz)$ for $n=0,2,4$ can be set to zero to reduce the problem
to an algebraic one and obtain the
solution. From the $n=4$ term , we find a consistency condition on the
equation
parameters:
$
\nu=-h\,\mu\,/s .
$
Thus $c^4\,=\,24s/\mu\,\,(\,=\,-24h/\nu)$.
From the other terms we find $g,c, etc$ and conditions we need to
impose as constraints on the equation parameters.
Thence we can easily find $\beta,\delta,\,g$ and $\Omega$
Note that , if $\epsilon=0$ the solution simplifies, as
then $g\,=\,\frac{\beta}{10s}$, so
\[
f\,=\,\sqrt{\frac{\beta}{5}}\,\,(\frac{6}{s\,\mu})^{\frac{1}{4}}
\,\,sech[\sqrt{\frac{\beta}{10s}}\,t].
\]

{\bf 3.2 Black soliton.} Here we take If we take $f=c\,g\, tanh(g\,t  )$,
then we see that
$
f''(t  )/[g^2\,f(t  )]  = 2\, \,[  tanh^2(gt  )-1],
$
and that
$
 f''''(t  )/[g^4\,f(t  )]
= 8\, \,[  tanh^2(gt  )-1]\,[  3\,tanh^2(gt  )-2].$
Again, on dividing  by $f$, it is clear that the coefficients of
$tanh^n(gt  )$ for $n=0,2,4$ can be set to zero to reduce the
problem,   and we then determine all the solution parameters .

{\bf 4.1. Chirped bright soliton.} This involves the function
$f\,=\,a(t  )\,exp[id\,log\, a(t  )]$,
where $a(t  )=g c \, sech(g\,t  )$.
With this form, the derivatives can still be written in a convenient way.
For example,
$
f''(t  )/[g^2\,f(t  )] = -\,  (d-i)\,[i+(d-2i) \,
tanh^2(gt  )].
$

We  need to find the roots of a 4th order polynomial in $d$:
\begin{equation}\label{poly4}
 (\mu\,h+s\,\nu)(d^4-35d^2+24)\,+\,10d(\nu\,h-s\,\mu)(5-d^2)\,=\,0.
\end{equation}
  Thus $d$ depends on a
balance of the highest order derivative (4th) and strongest
nonlinearity and is quite different from
that of the CGLE \cite{book}  or CGLE with an integral term
\cite{winter}  where $d$ is determined by $\beta$ and $\epsilon$.
Eqn(\ref{poly4}) provides insight into the chirpless ($d=0$)
case, as we see that $d=0$ is a root of eqn.(\ref{poly4}) when
$\mu\,h+s\,\nu=0$.

{\bf 4.2 Chirped black soliton} Here
$f(t  )\,=\,a(t  )\,exp[id\,log \,b(t  )],
$
where $a(t  )=r \,g\, tanh(g\,t  )$ and $b(t  )=g   \,
sech(g\,t  )$, and we proceed as above to find the solution.

{\bf 5. Energy and momentum balance.} As any other equation describing
dissipative systems, the quintic complex S-H equation does not have
any conserved quantities. Instead, we can write balance equations for
the energy and momentum. A study of the quintic complex S-H eqn., using the energy balance
approach (\cite{book},\cite{winter})  leads to the following
evolution equation for the energy:
\begin{equation}\label{en}
\frac{d}{dz  }\int\limits_{-\infty}^{\infty}|\psi|^2 dt =
2 \int\limits_{-\infty}^{\infty} \Big{[}\delta |\psi|^2-\beta
\Big{|}\frac{\partial \psi}{\partial t}\Big{|}^2 -\,s\,|\psi_{tt}|^2+
\epsilon |\psi|^4 + \mu
|\psi|^6\Big{]} dt.
\end{equation}
By definition, the right hand side of this equation is the rate of
change of the energy. The term with $s$ here is new in comparison with
similar equation for CGLE \cite{book}. The form of this term  reflects its
role as a higher-order
band-limited gain. For any stationary exact solution, we need the r.h.s. of
eqn.(\ref{en}),
and also the rate of change of the momentum, to be zero. This provides
a way of finding or checking  solutions. The balance equation is also
an important tool in studying the interaction between the pulses
\cite{book}.

{\bf Conclusion.} Solutions presented here are novel examples of exact
solutions which exist for the quintic complex S-H equation. As for the Ginzburg
-Landau equation \cite{book} they certainly do not cover the whole set of
possible
solutions. In fact, they only represent a small subset of a variety of
soliton-like solutions. Other solutions have to be studied numerically and
this may
take considerable amount of simulations. However, finding the exact
solutions
is an important step in analysing the laser system with more involved
spectral properties which is described by the quintic complex S-H equation.


\vspace{-1.0cm}
\begin{thebibliography}{99}
\small{
\bibitem{r4}
 V  J Sanchez-Morcillo et al. , Generalized complex Swift-Hohenberg
equation for optical
parametric oscillators,
  {\it
Phys. Rev. A },{\bf 56},3237 (1997) .
\vspace{-0.3cm}
\bibitem{r5}  J Lega et al., Swift-Hohenberg equation for lasers,
  {\it Phys. Rev. Lett.}, {\bf 73}, 2978
(1994).
\vspace{-0.3cm}
\bibitem{Weiss}
 C.O.Weiss et al.,Spatial solitons in nonlinear resonators, in {\it
Soltion-driven photonics
(eds. A.D. Boardman and A.P.Sukhorukov), Kluwer, 2001}, pp.169-210.
\vspace{-0.3cm}
%
\bibitem{SB98} H. Sakaguchi and H. R. Brand, Physica D {\bf 117} 95
	(1998).
\vspace{-0.3cm}
\bibitem{book} N.~Akhmediev and A. Ankiewicz,
{\it `Solitons, nonlinear pulses and beams'},
Chapman \& Hall, London (1997) 336 pp.
\vspace{-0.3cm}

\bibitem{winter}
A. Ankiewicz, N. Akhmediev and P.Winternitz, Singularity
analysis, balance equations and soliton solution of the nonlocal
complex Ginzburg-Landau equation, {\it Jnl. of Enrg. Maths}
{\bf 36},(1999) 11-24.
}
\end{thebibliography}
 \end{document}